\documentclass[]{agujournal}
   \usepackage{natbib}
  \usepackage{amsmath}
  \usepackage{cite}
  
  \usepackage{float}
  
  \usepackage{hyperref}
  \hypersetup{
      colorlinks=true,
      linkcolor=blue,
      citecolor=blue,      
      urlcolor=blue
  }
  
  \usepackage{etoolbox}
  \newskip\mfilskip
  \newcommand{\mfilbreak}{\vspace{\mfilskip}\penalty -200%
    \ifdim\lastskip<\mfilskip\vspace{-\lastskip}\else\vspace{-\mfilskip}\fi}
  \pretocmd{\section}{\mfilbreak}{}{}

  \draftfalse
  
  \journalname{Journal of Geophysical Research}

\begin{document}
  
  \title{Machine Learning Approach to Earthquake Rupture Dynamics}
  
  \authors{Sabber Ahamed\affil{1}, Eric G. Daub\affil{1, 2}}
  
  \affiliation{1}{Center for Earthquake Research and Information (CERI), University of Memphis, TN}

  \affiliation{2}{Alan Turing Institute, London, United Kingdom}
  \correspondingauthor{Sabber Ahamed}{sabbers@gmail.com}
  
  \begin{keypoints}
  \item Two machine learning algorithms are used to predict if an earthquake can break through a fault with a geometric heterogeneity.
  \item Models built from the algorithms can predict rupture propagation or arrest with more than 81$\%$ accuracy.
  \item Machine learning can identify the underlying complex data patterns that determine the physics of earthquake rupture propagation.
  \end{keypoints}
  
    \begin{abstract}
      Simulating dynamic rupture propagation is challenging due to the uncertainties involved in the underlying physics of fault slip, stress conditions, and frictional properties of the fault. 
      A trial and error approach is often used to determine the unknown parameters describing rupture, but running many simulations usually requires human review to determine how to adjust parameter values and is thus not very efficient. To reduce the computational cost and improve our ability to determine reasonable stress and friction parameters, we take advantage of the machine learning approach. We develop two models for earthquake rupture propagation using the artificial neural network (ANN) and the random forest (RF) algorithms to predict if a rupture can break a geometric heterogeneity on a fault. We train the models using a database of 1600 dynamic rupture simulations computed numerically. Fault geometry, stress conditions, and friction parameters vary in each simulation. 
      We cross-validate and test the predictive power of the models using an additional 400 simulated ruptures, respectively. Both RF and ANN models predict rupture propagation with more than 81$\%$ accuracy and model parameters can be used to infer the underlying factors most important for rupture propagation. Both of the models are computationally efficient such that the 400 testings require a fraction of a second, leading to potential applications of dynamic rupture that have previously not been possible due to the computational demands of physics-based rupture simulations.
      
      \end{abstract}
      
      \section{Introduction}
      Damage due to earthquakes poses a threat to humans worldwide. Seismic hazard analysis is used to estimate the possible ground motion at a given location in a given period based on historical earthquake data and decay of ground motion intensity with distance. However, current approaches to seismic hazard analysis are largely empirical and may not capture the full range of ground shaking in future large earthquakes due to a lack of sufficient historical geological data. This leads to large uncertainties in hazard estimates. Ideally, this lack of data can be mitigated by employing physical models that supplement existing data with additional scenario events that can quantify the expected variability of ground shaking to provide a robust estimate of hazard and risk.
      
      Earthquake faults are incredibly complex systems that span a vast range of length and time scales, making it challenging to construct physical models that resolve all of the relevant physics. Information on the state of stress of the faults can be obtained from past earthquake focal plane mechanisms,  geologic observations, or direct well-bores or drill holes~\citep{zoback2010reservoir}. However, these stress analysis techniques typically neither constrain the full stress tensor nor cover a broad range of locations and depths. Therefore, direct \textit{in situ} measurements of the stresses and displacements during rupture are rarely available. Even if the stress state and microscopic physics governing earthquake slip are known with certainty, multi-scale modeling of earthquakes poses a vast computational challenge due to the range of length and timescales involved. Due to these limitations, physical modeling is not routinely used directly in estimating earthquake hazard. However, physics-based approaches like dynamic earthquake rupture simulations are frequently suggested as a way that such physical constraints could be incorporated into seismic hazard estimation~\citep{harris1998introduction, harris1999dynamic, graves2011cybershake, olsen2009shakeout, de2009dynamic}.
      
      Dynamically simulating earthquake rupture is challenging due to uncertainty regarding the underlying physics of earthquake slip. The stress conditions and frictional properties of faults are not well constrained~\citep{Duan2006, peyrat2001dynamic, ripperger2008variability, kame2003effects} although they, together with fault geometry, control the rupture process and determine the dynamics of slip as well as the resulting ground motions. Because earthquake rupture is a highly nonlinear process, determining parameter values is often done by making simplifying assumptions or taking a trial and error approach, which usually incurs expensive computational and numerical costs~\citep{douilly20153d, ripperger2008variability, peyrat2001dynamic}. Therefore, this limits the applicability of the simulations, as due to the high computational expense they cannot be easily integrated into other calculations such as inversions or seismic hazard analysis. 
      
      While rupture simulations compute the slip and ground motions with a high level of detail, seismic hazard analysis usually only requires more general characteristics of an earthquake such as moment magnitude or peak ground velocity~\citep{Petersen2014}.
      Thus, rupture simulations may only need to approximate such characteristics to be useful in the hazard analysis. Machine learning is a promising approach for reducing the computational cost involved in estimating such attributes of complex data sets. Many recent studies show that a machine learning approach can be beneficial in a variety of applications including seismic event detection, hazard analysis, and fault detection from unprocessed seismic data.
      For example,~\citet{RouetEartqhaukeMachine} used a random forest-based algorithm to identify signals from the laboratory-generated acoustic measurements with predictive power. Their model was able to accurately forecast future failure events using only a window of data much shorter than the time between events. ~\citet{perol2018convolutional} developed a convolutional neural network algorithm $\texttt{ConvNetQuake}$ to detect earthquakes. $\texttt{ConvNetQuake}$ showed the ability to identify 20 times more earthquakes accurately than were contained in the catalog listed by the Oklahoma Geological Survey.~\citet{last2016predicting} explored machine learning methods to predict the largest possible magnitude of an earthquake likely to occur next year based on the previously recorded seismic events in Israel and its neighboring countries. The model was able to predict the maximum earthquake magnitude with 69.8$\%$ AUC (Area Under the Curve) accuracy.~\citet{araya2017automated} developed a neural network based model that can identify faults from raw seismic data bypassing the expensive multistep seismic processing, which is a significant promise in hydrocarbon exploration. These models illustrate the variety of ways that machine learning can take advantage of geophysical data to help us understand the complex structure and dynamics of the earth's crust. Another recent application of machine learning algorithm is to predict ground motion.~\citet{paolucci2018broadband} used a neural network to predict broadband earthquake ground motions from 3D physics-based numerical simulations. The authors showed that their algorithm could reproduce the high-frequency shaking based on the low-frequency simulation results, illustrating how physics-based models might be integrated into hazard modeling.
      
      In this paper, we describe a workflow for using machine learning to predict if an earthquake can break a fault with geometric heterogeneity, a complex problem that is dependent on geometry, stress, and fault friction coupled to elastic wave propagation. We start by presenting the data generated from dynamic rupture simulations and normalizing the data (zero mean and unit standard deviation) for training models. Then we examine two models built from the random forest and neural network and trained by a large number of earthquake rupture simulations. Finally, we discuss the potential to use machine learning models in predicting real surface fault rupture and estimating seismic hazard.

      \section{Rupture simulations and data preprocessing}
      \begin{figure}[ht]
        \begin{center}
        \includegraphics[scale=0.8]{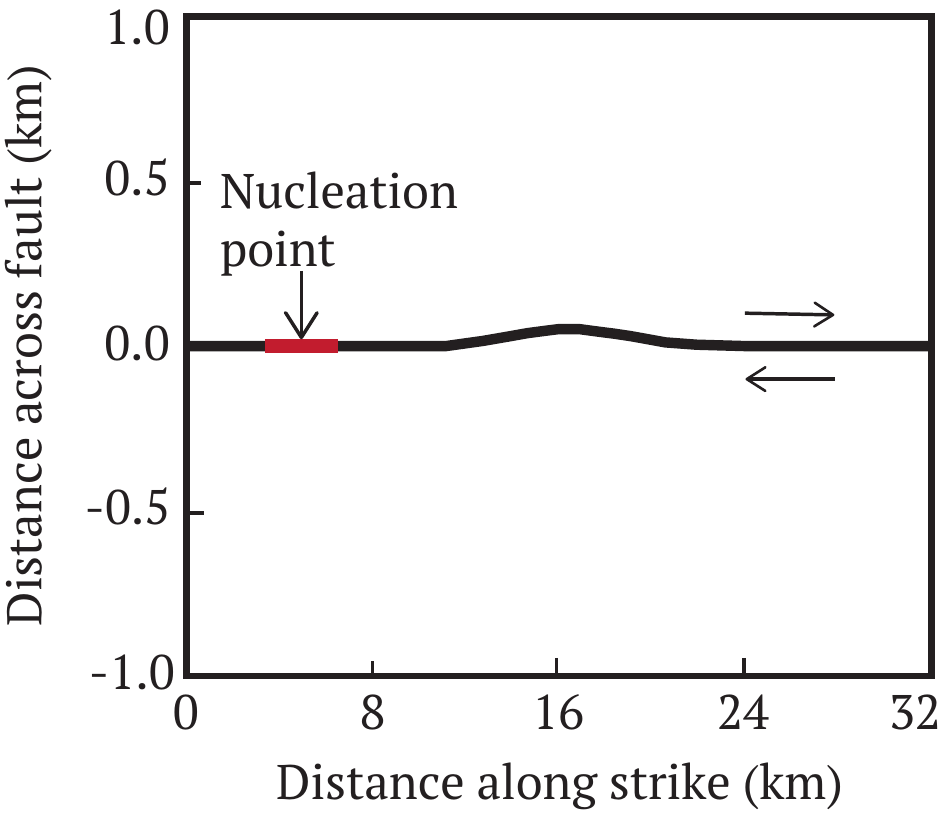}
        \end{center}
        \caption{The schematic diagram showing the fault geometry for the preliminary set of rupture models (vertical scale exaggerated). The domain is 32 km in length along the strike of the fault and 24 kilometers wide across the fault. The diagram shows a zoomed view of the fault for better visualization of the barrier. Rupture initiates at the nucleation patch (red) which is ten kilometers from the curved geometric barrier. The half-width and height of the barrier, stress tensor and friction parameters are varied randomly over the simulations.}
        \label{fig:rupture_domain}
        \end{figure}
      
      We produce a set of 2,000 rupture simulations based on the geometry illustrated in Fig.~\ref{fig:rupture_domain} using $\texttt{fdfault}$~\citep{fdfaultEricGithub} a finite difference code for numerical simulation of elastodynamic fracture and friction problems. A dynamic rupture model solves the elastodynamic wave equation coupled to a friction law describing the failure process. The simulation dynamically determines fault slip based on the initial stress conditions, the elastodynamic wave equations, and frictional failure on that fault.
  
\begin{figure}[ht]
          \begin{center}
          \includegraphics[scale=0.8]{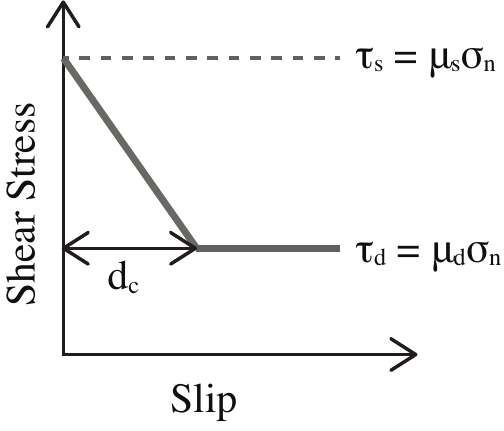}
          \end{center}
        \caption{Slip-weakening friction for an earthquake fault. The fault begins to slip when the shear stress reaches or exceeds the peak strength $\tau_s$. Over a critical slip distance $d_c$, $\tau_s$ decreases linearly to a constant dynamic sliding friction $\tau_d$. The shear strength is linearly proportional to the (possibly time-varying) normal stress $\sigma_n$, and the friction coefficient varies with slip between $\mu_s$ and $\mu_d$.}
          \label{fig:slip_weakening}
        \end{figure}

      
      The fault in the simulation is planar, with a Gaussian geometric heterogeneity at the center of the fault. Rupture is nucleated 10 km to the left of the barrier and propagates from the hypocenter towards the barrier. The rupture on the fault is governed by the linear slip-weakening law (\ref{fig:slip_weakening}). The fault starts to break when the shear stress ($\tau$) exceeds the peak strength $\tau_s = \mu_s \sigma_n$, where $\mu_s$ and $\sigma_n$ are the static friction coefficient and normal stress, respectively. Over a critical slip distance $d_c$, the friction coefficient reduces linearly to constant dynamic friction $\mu_d$.

    Fault geometry, stress state, and friction are all critical considerations for whether or not a rupture propagating from left to right can break the barrier. For right-lateral slip on the fault, the fault geometry is such that the closest side of the barrier is a restraining bend, which inhibits rupture, while the far side is a releasing bend that promotes rupture. Due to the fault geometry, all three stress components influence the shear and normal tractions on the restraining and releasing bends. For example, normal stress increases around the restraining bend depend on the bending angle while it decreases on the releasing bend, making the barrier a challenging effect to systematically analyze for rupture propagation. In this project, the half-width of the barrier varies between 1 and 2.1 km while the height is set in the range of 0-10$\%$ of its half-width. These values are large enough that the barrier has a non-negligible influence on the actual shear and normal tractions. Thus the combination of the stress state with the complexity of the fault geometry makes it difficult to predict the ability of the rupture to break the barrier due to these nonlinear effects.

\begin{figure}[ht]
      \begin{center}
      \includegraphics[scale=0.8]{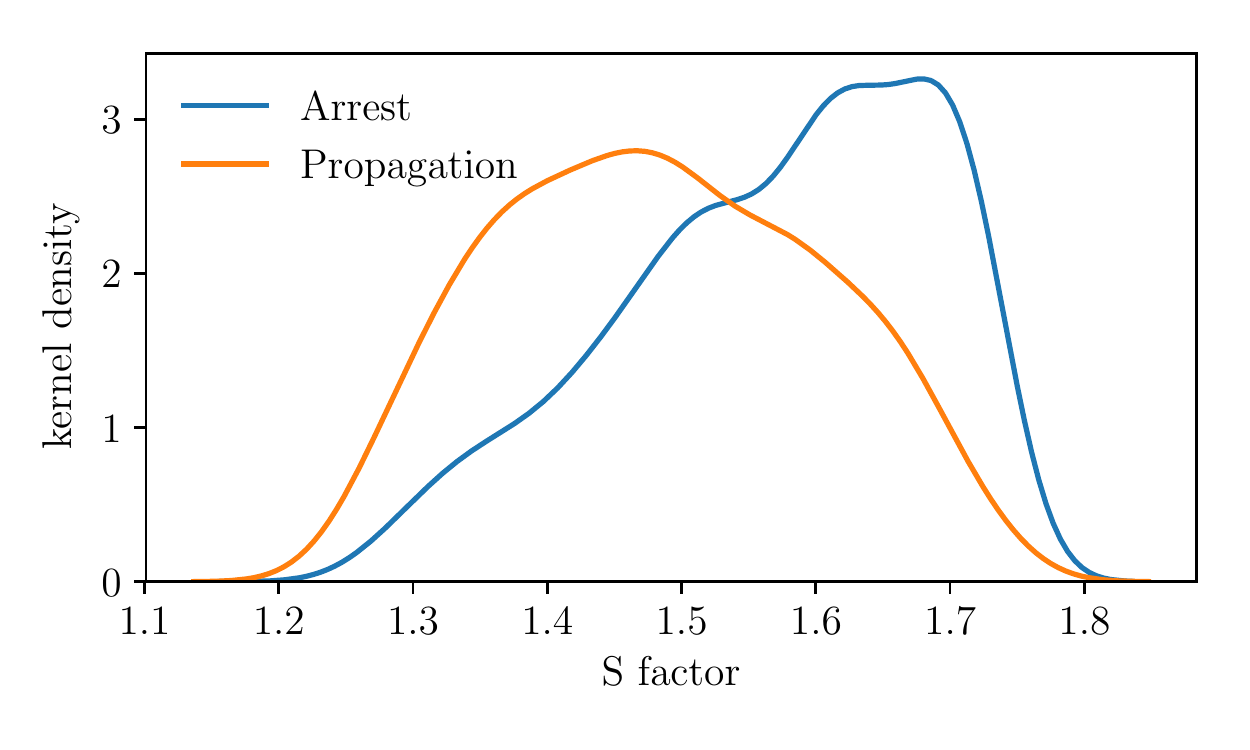}
      \end{center}
      \caption{Smoothed probability density function of $S$-ratio for rupture arrest and propagation. Ruptures that arrest typically have a large $S$-ratio and a smaller value indicates that rupture is close to failure. The magnitudes of $S$-ratio for rupture arrest and propagation have a similar range and have a broad overlaping region that indicates that the S-ratio has limited power as a discriminant.}
      \label{fig:kde_s_factor}
    \end{figure}

    
    Rupture simulations frequently use the shear over normal stress or the $S$-ratio~\citep{andrews1976rupture, das1977numerical} to characterize the ability of a rupture to propagate on a fault. The $S$-ratio is calculated as $S = (\tau_s -\tau_i)/(\tau_i - \tau_d)$, where $\tau_i$ is the initial shear stress, and $\tau_s$ and $\tau_d$ are the peak and residual shear stresses, respectively (Fig.~\ref{fig:slip_weakening}). For this particular problem, however, we find that neither metric can reliably predict if the rupture can break the barrier for a given set of parameters. We find that the $S$-ratio does offer some utility as a discriminant for rupture, but only in the sense that large values of S are more likely to arrest. Smaller values of the $S$-ratio traditionally indicate a rupture that is closer to failure, but we find that for this particular problem small $S$-ratios still arrest in many situations. For example, the Fig.~\ref{fig:kde_s_factor} shows the smoothed probability density function of the $S$-ratio for rupture arrest and propagation estimated from the training set of rupture data. From the figure, it is obvious that rupture propagation and arrest have large overlapping regions that make the S-ratio a poor discriminant for this particular problem. This is because of the role of the fault geometry, combined with the out of plane normal stress component. 

    For simple planar faults with uniform stress and strength conditions, the S-ratio and shear over normal stress are indicative of the stress situation over the entire fault, and thus predictive of rupture over the whole fault. For a rough fault, on the other hand, the actual tractions on the barrier can be significantly different from those for a flat fault, and thus ruptures that would ordinarily propagate may arrest. Thus, while this might appear at the outset to be a rather simple rupture problem, it captures the essential physics of rupture propagation on more complex fault geometries and provides a useful test case. Therefore, it allows us to determine if a machine learning algorithm can predict the ability of a rupture to propagate under a given set of conditions. The 2000 simulations are divided into training (1600) and test (400) sets. 1600 training datasets are used to train and validate the model while the rest of the 400 test simulations are used to test the performance of the models. We control the following eight parameters to vary the rupture behavior in each simulation:

      \begin{enumerate}
        \item The geometric barrier half-width follows a uniform distribution between 1 and 2.1 km.
        \item The barrier height follows a uniform distribution between 0 and 10$\%$ of the barrier half-width.
        \item The fault normal stress (in-plane, or IP) follows a uniform distribution between -10 and -160 MPa (negative indicates compression hence smaller means higher stress). Note that while this sets the normal stress on the flat part of the fault, while the normal stress on the barrier varies spatially due to the geometry.
        \item The other normal stress component, which we term the out-of-plane (OP) normal stress, follows a uniform distribution between $\pm$25$\%$ of the fault normal stress.
        \item The dynamic friction coefficient ($\mu_d$) follows a uniform distribution between 0.2 and 0.6.
        \item The friction drop (static minus dynamic friction) follows a uniform distribution between
        0.2 and $0.8-\mu_d$.
        \item The shear stress is set to be proportional to the normal stress. The proportionality coefficient is the dynamic friction coefficient plus 2$\%$ of the friction drop ($\mu_s-\mu_d$), plus a random number times 13$\%$ of the friction drop. In other words, the initial stress is chosen to be in a relatively narrow range of values where it is not trivial to predict whether or not the rupture will be able to propagate. Stress is always higher than the dynamic strength, but never very close to the static strength. This small range reflects the idea that earthquakes occur on faults once the shear stress on the fault reaches the minimum stress needed to rupture the entire fault, so such values should be representative of realistic values of the initial shear stress on natural faults.
        \item The slip weakening distance ($d_c$) follows a normal distribution centered at 0.4 m with a standard deviation of 0.05 m.
        \end{enumerate}

      \begin{table}[ht]
        \caption{Five sample training examples before preprocessing}
        \label{tab:sample_datasets}
        \centering
        \begin{tabular}{p{2.40em} p{2.40em} p{4.0em} p{3.0em} p{3.75em} p{3em} p{2em} c c}
        \hline
        Height (Km) & Half-width (Km) & OP Stress (MPa) & Shear stress (MPa) & IP Stress (MPa) & Fric. drop & Dyn fric. & $d_c$ (m)& Output\\
        \hline
        0.104 & 1.146 & -102.509 & 58.619 & -117.766 & 0.484 & 0.217& 0.296& 0 \\
        0.088 & 1.304 & -136.062 & 51.391 & -126.715 & 0.346 & 0.448& 0.406& 1 \\
        0.099 & 1.260 & -117.559 & 40.972 & -115.529 & 0.293 & 0.502& 0.389& 1 \\
        0.116 & 1.191 & -128.169 & 94.021 & -157.830 & 0.572 & 0.203& 0.409& 0 \\
        0.018 & 1.108 & -106.350 & 29.149 & -101.379 & 0.253 & 0.325& 0.398& 1 \\[1ex]
        \hline
        \end{tabular}
      \end{table}
      
      Five sample rupture examples before preprocessing are listed in Table~\ref{tab:sample_datasets} and shows the different parameter values with the corresponding output, 0 or 1. An output value of 1 means that the rupture propagated through the restraining geometric barrier while 0 indicates that the rupture arrested before reaching the center of the barrier.  All the parameters of the dataset are normalized to have zero mean and unit standard deviation. Note that this puts the problem in the form of a standard classification problem, for which many algorithms have been developed~\citep{carbonell1983overview, michie1994machine}.
      
      \section{Classification Stategy}
      We use two different classification algorithms to create predictive models to determine if a rupture can break the fault in our model. One is the random forest decision tree algorithm~\citep{breiman2001random} and the other is the artificial neural network~\citep{rosenblatt1958perceptron,  rumelhart1988learning}. These two algorithms have been selected due to their flexibility in handling a range of classification problems with a large feature space as well as the fact that their estimated parameter values can provide insight into the underlying dynamics. Another important aspect is that, while many learning algorithms such as nearest neighbor algorithms may work well for a particular problem, they may not necessarily tell us much about the underlying physics. 
  
      Since our training dataset is imbalanced (65$\%$ of ruptures arrest while 35$\%$ of ruptures propagate), cost-sensitive learning~\citep{chen2004using} is beneficial in many situations to make our models more suitable to learn from imbalanced data. Therefore, imposing a strong penalty on misclassification of minority class can improve performance. We assign a high weight (i.e., higher misclassification cost) to the minority class (rupture propagation) using the `balanced` class strategy, where the class weights are given by the number of total samples / (number of classes $\times$ number of samples in each class). The above formula is used to calculate the weight for rupture arrest and propagation classes, which are  0.768 and 1.431 respectively. These parameters are used to weight the training examples in computing the cost function, which is then minimized when training the model.
      
      \section{Evalution Metrics}
      Since the data set is imbalanced, we use recall, precision, F1 score, and accuracy as the evaluation criteria. Accuracy is not always a good indicator of model performance in these cases, as poor performance on the rare class can be masked by good performance on the much larger numbers of the more common class. Recall measures the proportion of the true predicted positive examples among the total positive cases. The recall is a metric to evaluate the model performance on all positive samples. Precision, on the other hand, measures the proportion of the true positive examples out of the total number of predicted positives. Precision helps us to analyze the model quality regarding the positive predictions. The F1 score is an evaluation metric that combines precision and recall by taking the harmonic mean of them. The F1 score is an auxiliary evaluation metric that combines precision and recall in a way where the model must exhibit high precision and recall to obtain a good F1 score. F1 score is calculated as: 2 $\times$ (precision $\times$ recall) / (precision + recall).
      
      We also examine the confusion matrix to evaluate model performance. The matrix gives detailed information on true positive (TP), true negative (TN), false positive (FP), and false negative (FN) predictions. TP and TN are the examples where the rupture propagates or arrests, respectively, and the models predict them correctly. The FPs are situations where a rupture was predicted to propagate but actually arrested, and the FNs are examples where an arrest was predicted, but the rupture propagated. All of the metrics provide us with a better understanding of model performance.
      
      \section{Random forest classifier}
      
      Ensemble machine learning algorithms have been widely used because of their excellent accuracy, robustness, and ease of use. The method combines multiple independent learning algorithms (weak learners) to achieve better predictive performance than could be obtained from any of the single learners alone~\citep{opitz1999popular, polikar2006ensemble, rokach2010ensemble}. The ensemble makes a final prediction based on a majority voting from the individual components of the group. There are two types of ensembles commonly used: bagging and boosting. Bagging~\citep{breiman1996bagging} involves having each learner in the ensemble vote with equal weight while in boosting ~\citep{dietterich2000experimental}, the prediction is made by taking a weighted vote of the predictions. Weights are proportional to each learner's accuracy on its training set. Although the ensemble methods can often perform better than a single learner, it requires increased storage, extensive computation, and has a complex structure to interpret due to the involvement of multiple classifiers in decision making. 
      
      \begin{figure}[ht]
        \begin{center}
        \includegraphics[scale=0.8]{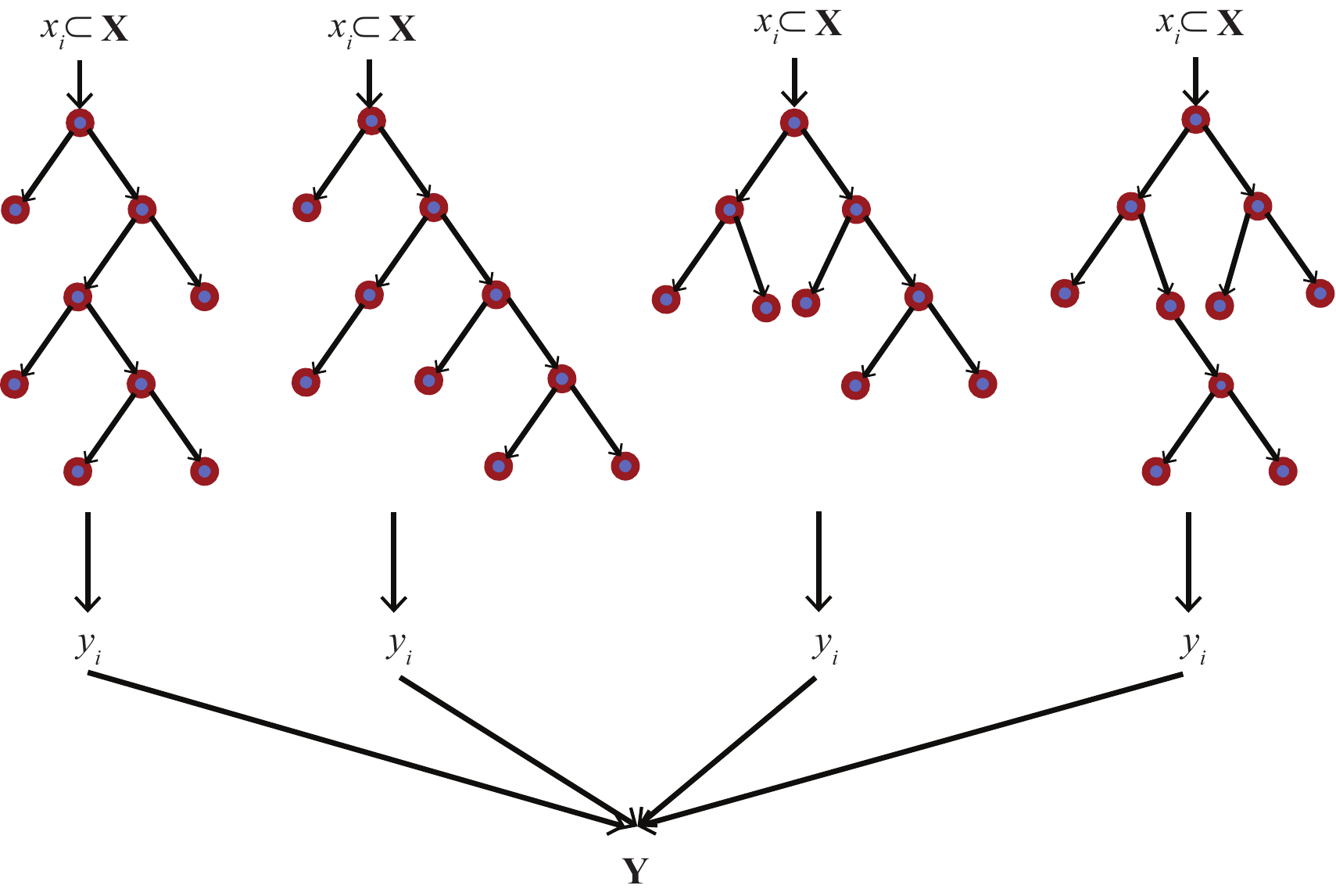}
        \end{center}
        \caption{The schematic diagram shows the ensemble of multiple decision trees known as the random forest. In each tree, the root node is at the top, and each internal node separates the data based on a decision using one of the input variables. An edge connects child nodes with a parent node. We use the ID3 algorithm~\citep{quinlan1986induction} to generate a single decision tree. In each tree, the input parameters ($\textit{x}_i$) are the subset of complete training data ($\textbf{X}$) and have subset features which are selected by random sampling with replacement. The random forest is the combination of multiple decision trees combined to make the final predicted class ($\textbf{Y}$).}
        \label{fig:random_forest_diagram}
      \end{figure}
      
      A fast algorithm such as decision trees is commonly used in ensemble methods to reduce the computational cost.  Fig.~\ref{fig:random_forest_diagram} shows a schematic of the ensembles of multiple single decision trees. Random forest (RF) is a bagging type of ensemble classifier~\citep{breiman2001random} that uses many single trees to make predictions. RF can be used for both classification and regression. In this project, we use the ID3 algorithm~\citep{quinlan1986induction} that uses information gain as the splitting criteria for a single decision tree.
      
      For the random forest classification, we use the $\texttt{scikit-learn}$~\citep{scikit-learn} library for implementing the algorithm in \texttt{Python}. Three parameters are optimized to improve the model performance: (1) maximum depth:  the maximum depth of the tree (2) minimum samples split: the minimum number of samples required to split an internal node and (3) number of estimators: the number of trees in the forest. To find the best parameters, we perform a grid search over the specified parameter values using the cross-validation technique to assess model performance. The best parameter value of maximum depth is 10, minimum samples split is 40,  and the number of estimators is 20. Cross-validation involves dividing the training data into two parts: the training data is used to estimate the model parameters and the validation data to determine how well the model generalizes to unseen data. By optimizing the model performance on the cross-validation data, we select the values of these model parameters.

      \subsection{Important features}
      \begin{figure}[ht]
        \begin{center}
        \includegraphics[scale=0.8]{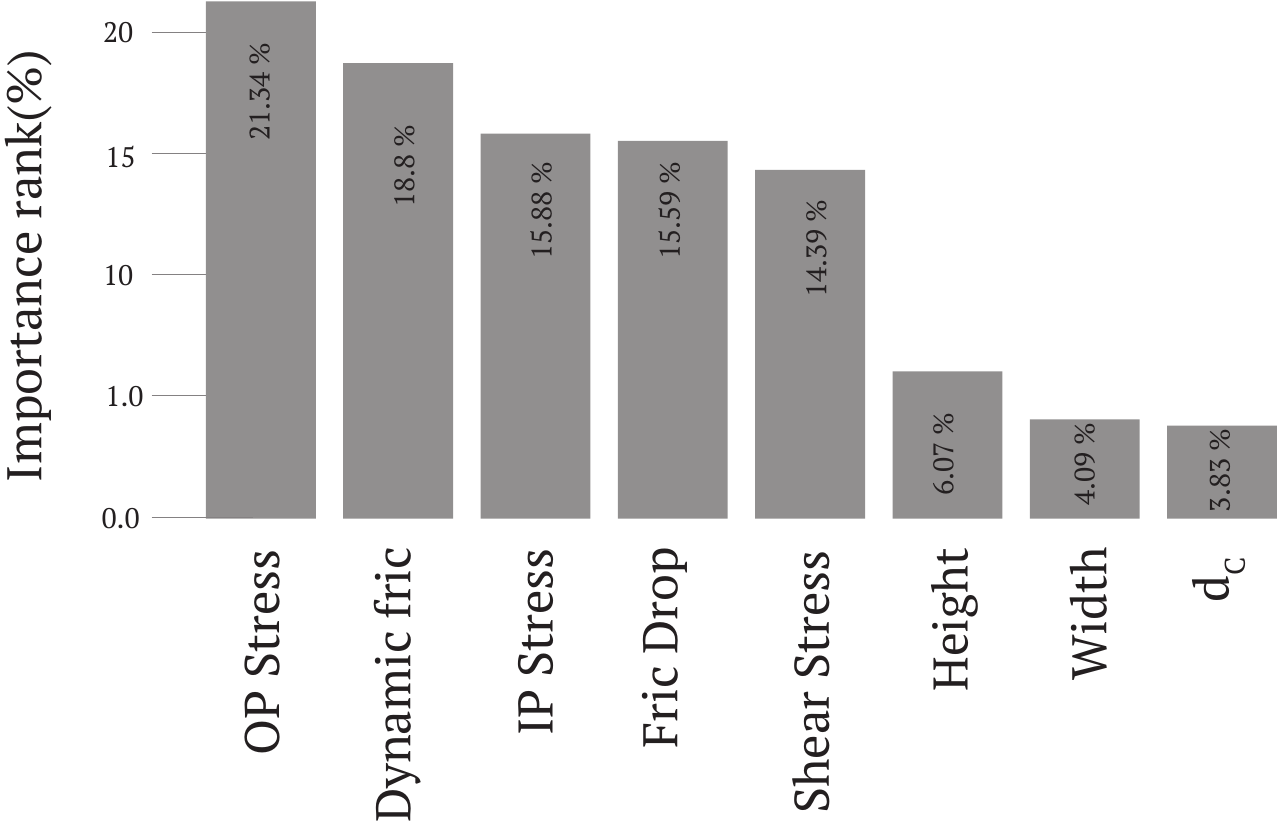}
        \end{center}
        \caption{The bar chart shows the relative importance (in $\%$) of the input features given by a random forest classifier.  OP normal stress has the highest influence (21.34 $\%$) on the decision followed by the dynamic friction coefficient (18.8 $\%$). IP normal stress, friction drop, and shear stress components show similar importance (15 $\%$). Geometric features such as height and half-width of the barrier and slip weakening distance ($d_c$) are identified as being less influential in making a decision. This bar chart and associated scripts, data are available under CC-BY ~\citet{Ahamed2017}.}
        \label{fig:important_features}
      \end{figure}
      
      
      The random forest algorithm allows the evaluation of the important features of a classification task. The most important features tend to occur higher in the decision trees, so we can see in this manner in which features are most predictive of rupture or arrest. Fig.~\ref{fig:important_features} shows the weighting of important features by percentage. Stress components, dynamic friction, and friction drop are the most important features to determine if an earthquake can break through a fault. These features account for 86$\%$ of the total predictive power. 
      This result is consistent with many studies that have shown that earthquake rupture initiation, its propagation, and termination are highly sensitive to stress and friction properties
       ~\citep{duan2007nonuniform, peyrat2001dynamic, ripperger2008variability, kame2003effects}.
      
    OP normal stress alone has the greatest (21.34$\%$) classification contribution followed by dynamic friction (18.80$\%$). IP normal and shear stress components and friction drop have almost equal importance (15$\%$). On the other hand, geometric features height and half-width have a less significant effect on the predictive power, indicating that the fault geometry is less important. This is likely because the stress and friction are important both with and without a geometrical heterogeneity on the fault, and regardless of how strong the heterogeneous fault geometry is, as long as it is loaded sufficiently close to failure it will still be able to propagate. Additionally, the influence of the OP normal stress is due entirely to the varying geometry, and the algorithm may find that the fault geometry varies less widely across the simulations, making the OP normal stress the variable that gives the most reliable predictions. The OP normal stress also varies more widely than the other stress parameters, so this may exhibit a stronger influence when constructing predictive models. This does not mean that fault geometry is unimportant, as it is well known that fault roughness influences the residual stress heterogeneities on the fault, which has a significant effect on earthquake recurrence probability~\citep{zielke2017fault}. Instead, this result suggests that if the geometry does not vary too much, the stress and frictional conditions are more useful to make predictions about the rupture process. Simulations using more complex fault geometries such as fractal faults may show a stronger sensitivity to features based on the fault geometry when compared to the simple barrier considered here.

      \subsection{Classification result}
      We use 400 test data to estimate the generalization error of the random forest model. Table~\ref{tab:rf_confusion matrix} shows the confusion matrix that contains information about actual and predicted classifications performed by the model. The RF classifier accurately predicts 218 test cases as TN. Similarly, 107 example ruptures are correctly identified as TP. On the other hand, 21 and 54 test cases were inaccurately classified as FP or FN, respectively.
      
      \begin{table}[ht]
        \caption{Confusion matrix given by the random forest model based on 400 testing data}
        \label{tab:rf_confusion matrix}
        \centering
        \begin{tabular}{l c c}
        \hline
         & Negative & Positive\\
        \hline
        Negative (Rupture arrested) & TN = 221 & FN = 51 \\
        Positive (Rupture propagated) & FP = 24 & TP = 104 \\
        \hline
        \end{tabular}
      \end{table}
      
      
      \begin{table}[ht]
        \caption{Random forest classification results based on 400 testing data}
        \label{tab:rf_classification_report}
        \centering
        \begin{tabular}{l c c c c}
        \hline
        Class & Precision & Recall & F1 score & Number\\
        \hline
        Negative (Rupture arrested) & 0.91 & 0.80 & 0.85 & 272 \\
        Positive (Rupture propagated) & 0.66 & 0.84 & 0.74 & 128\\
        Average/Total & 0.83 & 0.81 & 0.82 &400\\[1ex]
        \hline
        \end{tabular}
      \end{table}
      
      Table~\ref{tab:rf_classification_report} shows the classification result using three evaluation metrics. The high score of recall (0.80) for the positive class and the high precision (0.91) for the negative class indicates that the number of true positives (128) misclassified as negative is small (21). A slightly lower F1 score for the positive class indicates that the model performance of the rupture propagation class is not as good as the performance of the rupture arrest class. Adding more rupture propagation data may improve the positive class performance by helping the algorithm distinguish more subtle patterns in the dataset.
      
      \section{Artificial Neural Network}
      
      Artificial neural networks (ANN) are inspired by how neurons are connected in the brain~\citep{rosenblatt1958perceptron}. A neural network consists of several units interconnected and organized in layers. The individual units are also known as neurons.  The neurons perform a weighted sum of its input, so its output can fit complex functions by combining a large number of weights. A layer can be connected to an arbitrary number of further hidden layers of arbitrary size before being combined in the output layer.  Hidden layers introduce complexity to the model, and given sufficient data to train the model, these additional layers can improve performance. However, increasing the number of hidden layers does not always help. Additional hidden layers can lead to overfitting~\citep{hinton2012improving, lawrence2000overfitting, lawrence1997lessons} where the network will memorize the training data, but generalize poorly to new data. Therefore, selecting the number of layers and units in each layer is one of the challenges of constructing ANNs that perform well on complex data sets.
      
      \begin{figure}[ht]
        \begin{center}
        \includegraphics[scale=0.8]{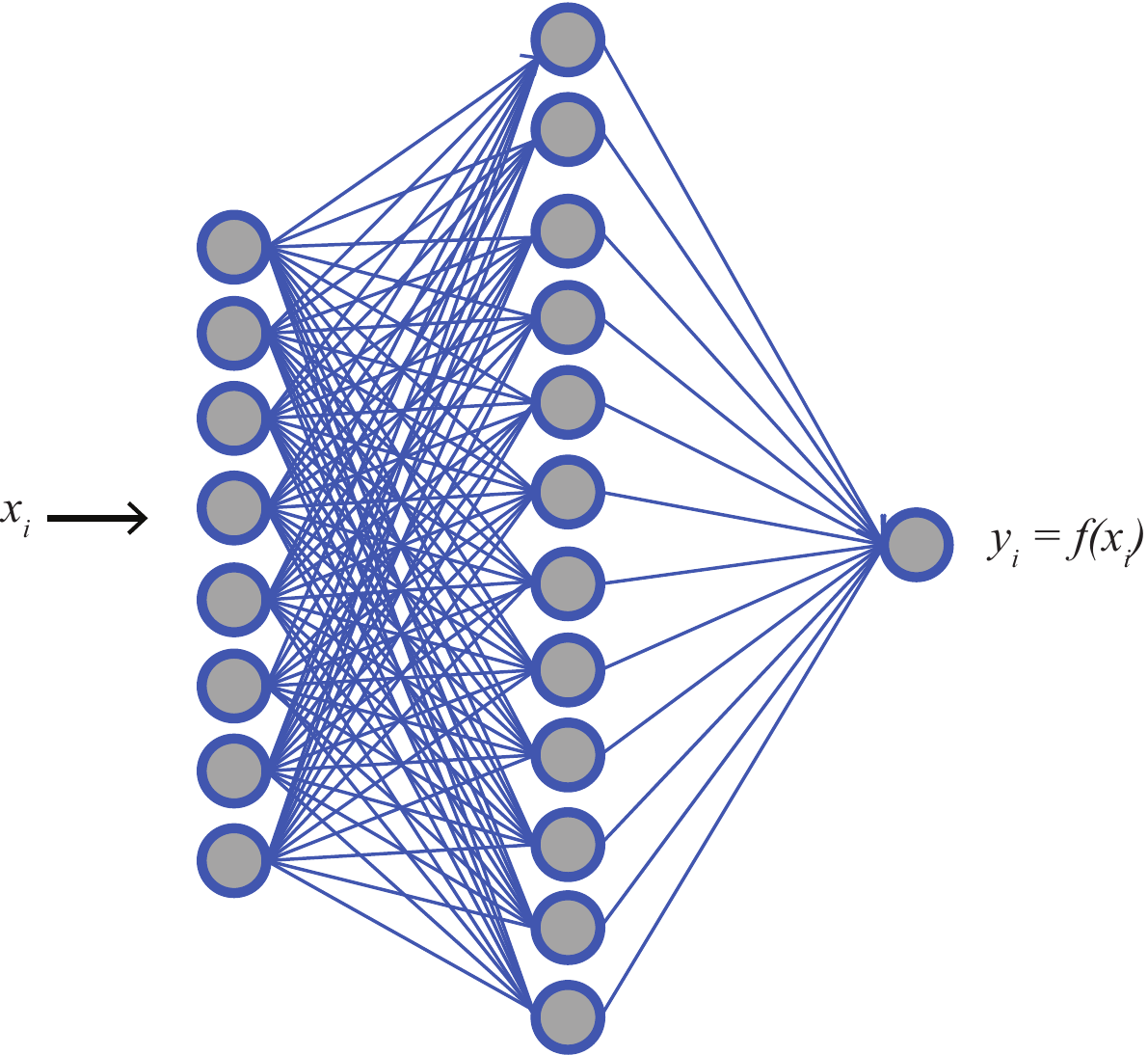}
        \end{center}
        \caption{Schematic diagram illustrating an artificial neural network. Normalized parameter values ($x_i$) are fed into the input layer, which are then combined into a hidden layer (center) using a set of parameter weights estimated from the training data. The resulting values from the hidden layer are then combined into an output layer, which computes a probability. Models may also contain multiple hidden layers, which enable the model to build more complex combinations of the input parameters.}
        \label{fig:nn_schematic_diagram}
      \end{figure}

      Figure.~\ref{fig:nn_schematic_diagram} illustrates the schematic diagram of the neural network topology we use in this work. The network has one hidden layer with 12 units. The eight input parameters are mapped to these 12 units, producing a $12 \times 8$  weight matrix for the model. As each input enters a unit, the output of the previous unit is multiplied by its weight. The unit then adds all these new inputs, which determines the output value of the intermediate unit. We then apply a nonlinear activation function ReLu ~\citep{hahnloser2000digital} to the output weight, which passes all the values greater than zero and set any negative output to be zero. Finally, the hidden layers combine the 12 outputs with the output layer and use the resulting weight to make predictions. We use \texttt{keras}~\citep{chollet2015keras}, a Python deep learning library, to build the model.

      In the ANN, 480 training examples (30$\%$ of the training set) are used for validating the model.
      To prevent overfitting (high variance, therefore, poor performance on unseen examples), we use several strategies: (1) cross-validation (a technique to assess model performance) (2) L-2 regularization that penalizes large weights and effectively reduces the variance of the model, hence prevents overfitting and (3) early stopping technique that stops the training when the difference between training and validation error start increasing rather than decreasing. In our case, if the validation accuracy does not improve in 20 consecutive training steps, the early stopping technique halts the training. 
      
      \subsection{Network parameters}
\begin{figure}[ht]
        \begin{center}
        \includegraphics[scale=0.6]{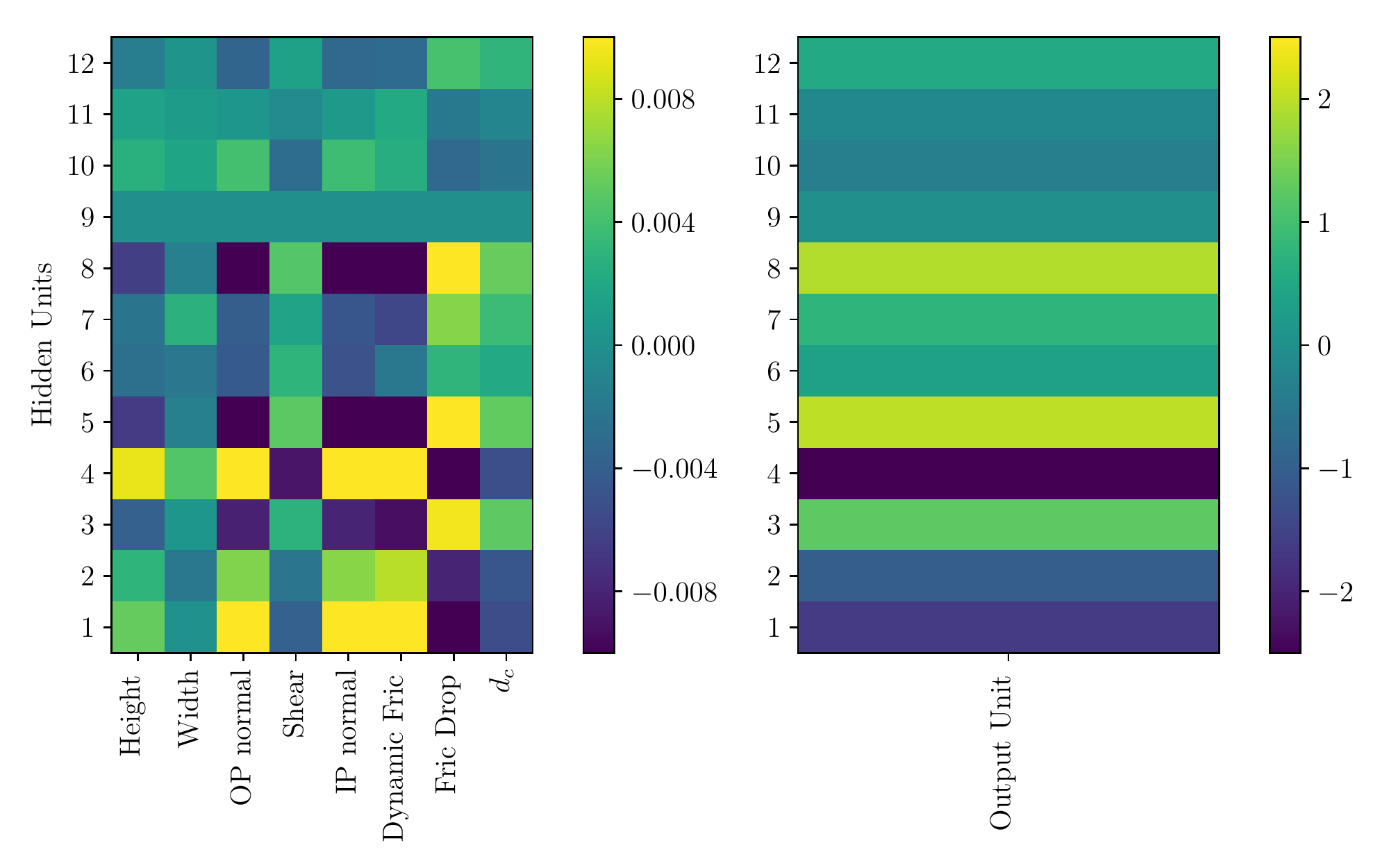}
        \end{center}
        \caption{The illustration shows the parameters learned by the ANN for the rupture model. The network has one hidden layer with twelve nodes. The left panel shows the weights that map the inputs to the hidden units. The eight input parameters are on the horizontal scale, and the twelve hidden units are on the vertical scale. All input parameters are normalized to have zero mean and unit standard deviation. The colors in each row indicate how the parameters are combined to form each hidden unit. The right panel shows the weights that are applied to each hidden unit to form the single output unit. A substantial positive value of the output unit indicates that the particular feature combination is predictive of propagation while a large negative weight of the output unit indicates that the combination is predictive of the arrest. These weights provide a physical understanding of the parameters selected by the neural network and give insight into the physics of rupture. This illustration and associated scripts are available under CC-BY ~\citet{Ahamed2017}.}
        \label{fig:nn_parameters}
      \end{figure}

      The weights learned by the neural network allow us to gain some insight into the combinations of input parameters that are most predictive of the ability of the rupture to propagate. To visualize the weights, we have constructed the weights versus neural units matrix plot. This is illustrated in Fig.~\ref{fig:nn_parameters}. The left panel shows the model weights mapping the eight inputs (horizontal) to the twelve hidden units (vertical). The right panel shows the weights that combine the hidden units into the one output unit on the right. The color scale indicates the range of weights. The weights in the model range from negative values to positive values which indicate inhibitory or excitatory influences.
      
      If the output layer weight is positive for a particular unit, then the parameter combination is a good predictor of rupture propagation, while a negative weight indicates the parameter combination is a good predictor of rupture arrest. Similarly, weights mapping the inputs to the hidden units that are positive indicators signify that large values of that input unit favor rupture. If the output unit has a negative weight, then large weights for a hidden unit indicate that large values of that parameter are predictive of the arrest. The weights provide insight into which parameter combinations are most predictive of rupture.
      
      All the weights related to height, half-width and slip-weakening distance $d_c$ have relatively small weights when compared to the stress and friction parameters (Fig.~\ref{fig:nn_parameters}). This means these three parameters have reduced the influence on the final prediction while the remaining five parameters have a much stronger contribution to the final predictions. We note the consistency with the random forest algorithm that gave the same importance rank for the parameters; that is, height, half-width, and $d_c$ are the least significant.

      
      Parameters illustrated in Fig.~\ref{fig:nn_parameters} also provide insights into the parameter combinations and their influence on determining the rupture propagation. Interestingly, we find that all of the units with large output might show similar patterns for the combined parameter values. For example, unit-4 has a large negative output weight. The unit has a large negative weight of shear stress, friction drop, and slip-weakening distance while height, half-width, OP, IP normal stress, dynamic friction coefficient, and friction drop have positive weight. This indicates that if a fault has low shear stress, and low friction drop, but high compressive OP,  IP normal stress, high dynamic friction coefficient, height, and half-width then it is likely that rupture would not propagate but arrest. On the other hand, unit-8 has a substantial positive output weight. The OP, IP normal and dynamic friction coefficient has a large negative weight while friction drop, slip-weakening distance, shear stress have high positive weight. Note that these are essentially the opposite values from those in unit-4, indicating a single underlying pattern in the data. Therefore, based on our input data the model has determined exactly how to best combine the various input parameters in a more sophisticated way than simply looking at shear over normal stress or the $S$ parameter. Additionally, the model provides a method for integrating the effect of the out of plane normal stress, which is known to be important for complex faults but is not accounted for quantitatively when examining the S-factor. 
      
  
      Our model is a simplification, and there is just one barrier with one shape and size, rather than many barriers with a much broader range of shapes and sizes. Although the fault geometry (half-width and height) has a relatively small influence in determining rupture, variations in barrier size still play an important role for local stress state on the fault surface. It is likely that when considering the full fractal fault, the fault geometry likely plays a more pronounced role as there are many more places where the geometry could cause the rupture to arrest. ~\citet{chester2000stress} used an analytical model of a wavy friction fault and found that the fault roughness has the most significant impact in determining the orientation and magnitude of principal stress. In a nonplanar fault geometry, narrower barriers have a more substantial stress perturbation in the sense that angle near the bend is sharper for the barriers of the same height, so the variation in traction at the releasing and restraining bend (Fig.~\ref{fig:rupture_domain}) is more prominent. Whereas if the barrier is broad, the angle around the bend changes gradually, and the stress perturbation at the restraining and releasing bend is less noticeable. Although the fault geometry is not as predictive as the stress and friction, the geometry still does play a role in understanding the rupture process, and the methods developed here give a straightforward way to account for them quantitatively.
      
      \begin{figure}[ht]
        \begin{center}
        \includegraphics[scale=0.8]{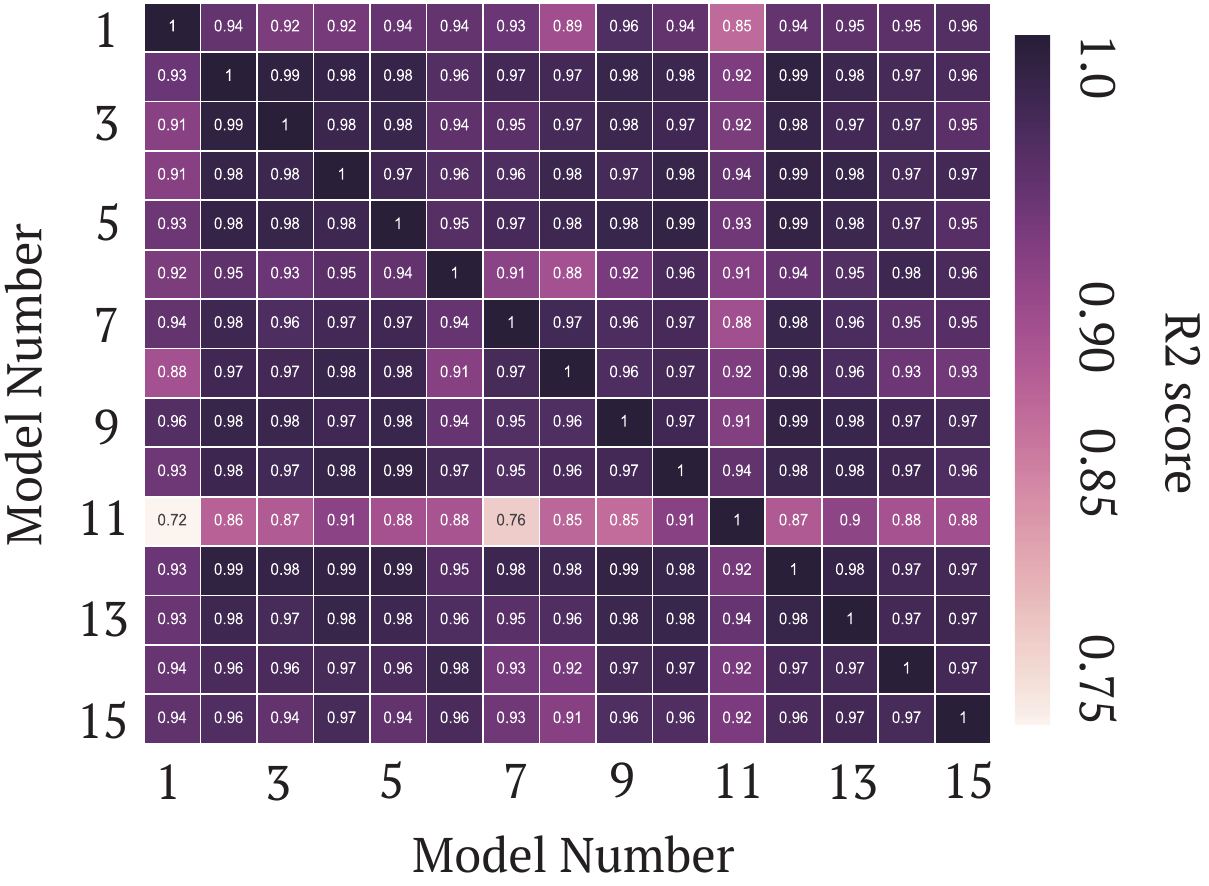}
        \end{center}
        \caption{The illustration shows the coefficient of determination (R2 score) among the weights learned by fifteen neural network models.  Weights are initialized by setting a random seed value for each different model. A single R2 score is calculated by first sorting the output weights, and then correlating the sorted weights with all other model realizations. Then we construct a correlation map from R2 scores of fifteen models to see if we obtain similar parameters independent of how the algorithm is initialized. Model-11 has the lowest correlation among all models, while the other models are strongly correlated. This indicates that the model picks up on robust features that are meaningful for understanding the rupture process. This illustration and associated scripts are available under CC-BY~\citet{Ahamed2017}.}
        \label{fig:correlation}
      \end{figure}
      
      We confirm that the parameter combinations found by the ANN approach are robust by repeating the fitting procedure. We build an additional 15 neural network models with the same training data set, but different initial weights to see if the models find the same features that are predictive of rupture. The average testing accuracy of the models is 83$\%$  which is very similar to the results shown in Table~\ref{tab:nn_classification_report}. Although the final weights vary slightly from model to model, they exhibit high correlation when we sort them in ascending order based on their output weight.  Fig.~\ref{fig:correlation} shows the determination of coefficient (R2 score) among the parameters (in ascending order) learned by the models.  Model-11 has the smallest correlation with the other model while models-10, and 12 have high correlations. Even though model-11 has the lowest correlation coefficient of 0.71 with model-1, it is high enough that it still contains similar patterns as the other models. The highly correlated weights indicate that the models are picking up on consistent features regardless of the random way that the model is initialized.
      
      \subsection{Classification result}
      
      \begin{table}[H]
        \caption{Confusion matrix given by neural network model on 400 testing data}
        \label{tab:nn_confusion matrix}
        \centering
        \begin{tabular}{l c c}
        \hline
         & Negative & Positive\\
        \hline
        Negative (Rupture arrested) & TN = 223 & FN = 49 \\
        Positive (Rupture propagated) & FP = 26 & TP = 102 \\
        \hline
        \end{tabular}
      \end{table}
      
      \begin{table}[h]
        \caption{Neural network classification results based on 400 testing data}
        \label{tab:nn_classification_report}
        \centering
        \begin{tabular}{l c c c c}
        \hline
        Class & Precision & Recall & F1 score & support\\
        \hline
        Negative (Rupture arrested) & 0.90 & 0.82 & 0.86 & 272 \\
        Positive (Rupture propagated) & 0.68 & 0.80 & 0.73 & 128\\
        Average/Total & 0.83 & 0.81 & 0.82 &400\\[1ex]
        \hline
        \end{tabular}
      \end{table}
      
      We use the same 400 test data to validate the ANN model. Table~\ref{tab:nn_confusion matrix} shows the confusion matrix (CM) that contains information about actual and predicted classifications. Interestingly the CM of the RF model is essentially the same as that of the ANN model, with two additional cases correctly predicted relative to the RF. The classification report (table~\ref{tab:nn_classification_report}) produced by the ANN model also shows nearly the same precision, recall, and F1 scores as the RF model. The overall testing accuracy of the model is $81.25\%$. This result suggests that several common classification algorithms are capable of accurately predicting the rupture results to the same level given the same input data. This also suggests that the misclassified test data are due to the number of training examples provided to the algorithms.

      \section{Misclassification analysis}
      
      \begin{figure}[ht]
      \begin{center}
      \includegraphics[scale=0.60]{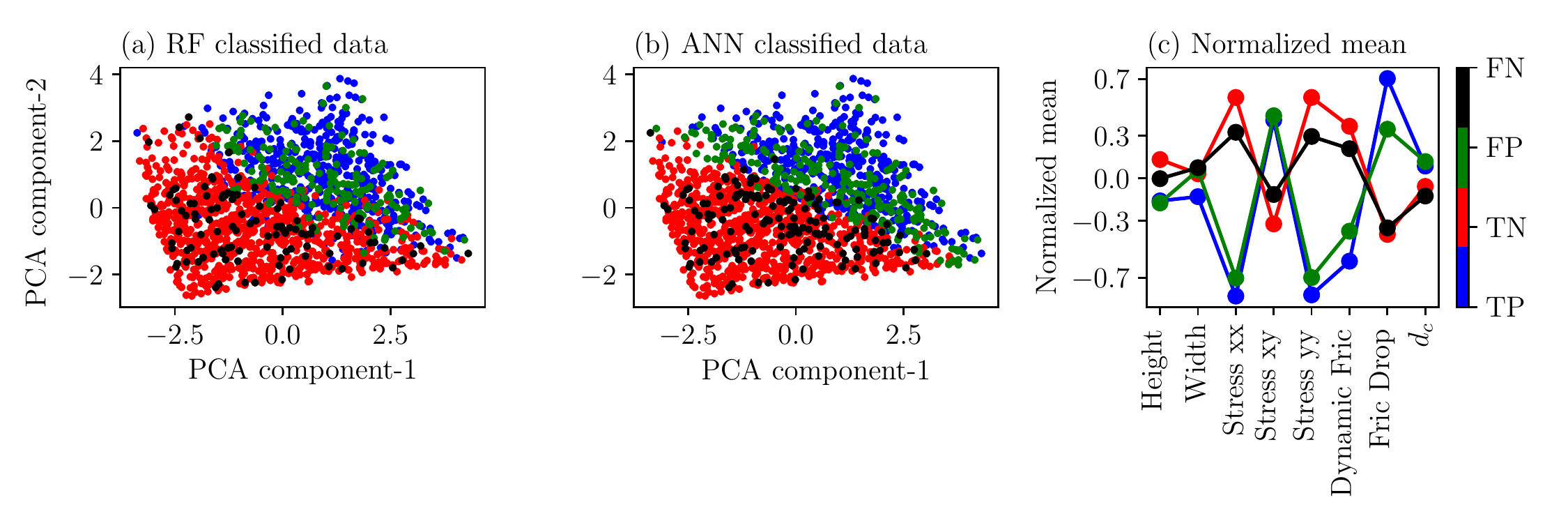}
      \end{center}
      \caption{The illustration shows the classified training data in two-dimensional space. We used Principal component analysis (PCA) to represent all the eight features in 2D space. PCA is a technique that uses Singular Value Decomposition to linearly reduce the data and project it to a lower dimensional space. (a) 2D representation (PCA-1 vs. PCA-2) of training data used for RF. Red and blue dots represent correctly predicted rupture arrest and propagation respectively while black and green dots are false negatives (FN) and positives (FP) respectively. (b) 2D representation (PCA-1 vs. PCA-2) of training data used for ANN. Symbols are the same as in Fig.~\ref{fig:misclassified_data} (a). The normalized mean value for each feature of the correctly and incorrectly classified data. In both of the RF and ANN models, rupture propagation and arrest are distinguishable based on the lower dimensional projection of the data. False positives are in the transition between true positive and true negative. When projected into the 2-dimensional space, some of the false positives overlap with the bulk of the true positives, while most of the false negatives are located in the same region of parameter space as most ruptures that arrest. This illustration and associated scripts are available under CC-BY~\citet{Ahamed2017}.}
      \label{fig:misclassified_data}
  \end{figure}
  
  Although the models generally provide robust predictions on the data, it performs less well on a number of false positives where rupture was predicted to propagate but arrested instead. To understand the model performance on the false positives and false negatives, we reduced all the training data into 2D using Principal component analysis (PCA). PCA is mathematically defined as an orthogonal linear transformation that transforms multidimensional data (eight dimensions in our case) to a lower dimension. Fig.~\ref{fig:misclassified_data}(a) and (b) show the scatter plot of the reduced training data set. The horizontal and vertical axes represent PCA component-1 and -2, respectively. Most of the rupture propagation (true positives) and arrest (true negatives) data are distinguishable in the PCA plots for the RF (Fig.\ref{fig:misclassified_data} (a) and ANN (Fig.\ref{fig:misclassified_data} (b).
  Fig.~\ref{fig:misclassified_data}(c) shows the normalized mean value of each feature of TP, FP, TN, and FN. This gives some insight that the misclassified examples tend to have parameter values that are on the outside edge of the averages of those that are correctly predicted. False positives tend to be located in the transition between true positives and true negatives. Some of the FP also overlap with true positives. Interestingly, the false positives have a similar mean value of all features when compared to the true positives illustrated in Fig.~\ref{fig:misclassified_data}(c). Likewise, the false negatives have similar mean feature values when compared to the true negatives. Most of the false negatives are located in the rupture arrest region. These examples likely were misclassified because we have insufficient data, and adding more data (especially more ruptures that propagated) could improve model performance. Another possible way to solve the problem is to use the Bayesian neural network (BNN)~\citep{gal2016uncertainty, 2018arXiv180107710M}. Future efforts will focus on ways to improve model performance on the misclassified examples.
  
  \section{Discussion}
  In this study, we develop machine learning models using the random forest and artificial neural network algorithms to predict earthquake rupture on a geometrically complex fault. The models provide a robust way to learn the parameter combinations responsible for rupture propagation. Because of the complicated fault geometry, nonlinear rupture process, and unknown material properties, it is difficult to predict if an earthquake can break through a fault~\citep{ohnaka1986dynamic, kanamori1993determination, marone1998laboratory, abercrombie2005can, warren2006systematic}. Moreover, discontinuous particle velocities across fault zones and tractions acting on the fault are governed by nonlinear friction laws, and obtaining these parameters in situ is challenging~\citep{mcgarr1978state, saffer2003comparison, saffer2001laboratory, kohlstedt1995strength}. Therefore, modeling earthquake rupture with a heterogeneous fault surface and unknown material properties remains a challenging computational problem~\citep{douilly20153d, ripperger2008variability, peyrat2001dynamic}. Our machine learning approach, on the other hand, provides a method to understand the rupture physics on complex fault and efficiently predict the result with high accuracy if a sufficient number of examples are provided.
  
  Classification results produced by the ANN and RF models show that they can capture most of the underlying patterns of rupture propagation. Interestingly, both models have around 84$\%$ recall on the positive class (rupture propagation), meaning that both of the models successfully learned most of the underlying complex data patterns responsible for rupture propagation.  Another exciting aspect of the ANN model is that it consistently finds the same hidden data patterns despite the various weight initializations. These underlying complex data patterns reflect the physics of the rupture propagation. For example, if a fault with geometrical heterogeneity has higher out-of-plane normal and shear stress and low static and dynamic friction, then it is likely that an earthquake rupture can break through the fault, and our model provides a way to evaluate this quantitatively.
  
  These models are highly efficient in determining whether a rupture is going to break through a fault. Both of the models take a fraction of a second to predict if a rupture can propagate given eight input parameters. This is a significant improvement over running a full rupture simulation, which takes about two hours of wall clock time on eight processors. Although the model performance in predicting earthquake rupture outperforms the traditional approach such as the S-parameter~\citep{andrews1976rupture, das1977numerical} or nondimensional prestress~\citep{bruhat2016rupture, kaneko2010supershear}, this approach is not meant to replace all rupture simulations, but rather to help determine parameter values. This approach might be combined with information such as a historical fault database and paleoseismic rupture areas to choose a parameter range consistent with past earthquakes in a region. The model predictions might also be incorporated into other complex calculations such as inversions or probabilistic seismic hazard analysis (PSHA). Since the models are computationally efficient, we can find a range of parameters likely to give a specific magnitude, which can be used to run some specific scenario events for estimating ground motions.
  
  The method could also be adapted into a probabilistic model for earthquake size given a few physically relevant parameters that can account for epistemic uncertainty by robustly considering different parameter value choices. We plan to expand our results to more extensive data sets as well as more realistic geometries drawn from fault databases, and develop methods to use machine learning to generate rupture models that are based on physics and can be used in PSHA.
      
  \section{Conclusion}
  We use the random forest and artificial neural network algorithms to predict if a rupture can propagate through a fault with a geometric heterogeneity. We first generate 2000 dynamic earthquake rupture simulations varying stress, friction parameters and height, and half-width of the restraining bend of the fault. In both of the models, 400 rupture examples are used to test the model performance. 
  Both of the models can consistently predict rupture with more than 81$\%$ accuracy. Data patterns identified by the models reflect the physics of the rupture propagation, and these patterns are robustly identified independently of how the models are initialized.
  
  Computationally, the models are highly efficient. Once the training simulations are computed, and the machine learning algorithms are trained, the models can make a prediction within a fraction of a second. This has the potential to allow for the results of dynamic rupture simulations to be incorporated into other complex calculations such as inversions or probabilistic seismic hazard analysis, something that would not be ordinarily possible. The method can also be applied to other complex rupture problems such as branching faults, fault stepovers, and other complex heterogeneities where the physics of earthquake rupture propagation is not fully understood. Machine learning provides a new way of approaching this complex, nonlinear problem, and helps scientists understand how the underlying geophysical parameters are related to the resulting slip and ground motions, thus helping us constrain future seismic hazard and risk.
  
\section{Acknowledgement}
This research was supported by the Southern California Earthquake Center (Contribution No. 7759). SCEC is funded by NSF Cooperative Agreement EAR-1033462 $\&$ USGS Cooperative Agreement G12AC20038. Also, thank CERI and University of Memphis HPC for providing support and computational resources. The source code for all graphs, plots, and models can be found in the Github repository: \url{https://github.com/msahamed/machine_learning_earthquake_rupture}.

\bibliography{machine_learning}
\end{document}